\documentclass{article}\usepackage{amssymb}
\newcommand{\ec}[1]{(\ref{#1})}
   \def\alp{\alpha}          \def\gam{\gamma}
   \def\Gam{\Gamma}   \def\del{\delta}      
   \def\eps{\epsilon}  
            
             \def\lam{\lambda}
           
   \def\sig{\sigma}   \def\Sig{\Sigma}      
        
   \def\ome{\omega}   \def\Ome{\Omega}
   \def\beq{\begin{equation}}\def\eeq{\end{equation}}
   \def\barr{\begin{eqnarray}}\def\earr{\end{eqnarray}}
   \def\basn{\begin{eqnarray*}}\def\easn{\end{eqnarray*}}
   \newcommand{\h}[1]{\hspace{2mm}#1}
   \def\nn{\nonumber}\def\pal{\partial}
   
   \newcommand{\mc}[1]{\mathcal{#1}}
   \newcommand{\parent}[1]{\left(#1\right)}
   \newcommand{\corchet}[1]{\left[#1\right]}
   \newcommand{\llaves}[1]{\left\{#1\right\}}
   
   \def\1/2{\frac{1}{2}}\def\im{\mathrm{i}}\def\wed{\wedge}

\begin{document}
\fontsize{12pt}{16pt}\selectfont
\begin{titlepage}
\title{Minimal Immersions and the Spectrum of Supermembranes}
\author{J. Bellor\'{\i}n\thanks{{\tt bellorin@delta.ft.uam.es}}
\hspace{3mm} and \hspace{3mm} A. Restuccia\thanks{{\tt arestu@usb.ve}} \\
{\fontsize{10pt}{14.4pt} \textit{Department of Physics,
Universidad Sim\'on Bol\'{\i}var, Caracas, Venezuela}}}
\date{}\maketitle

\begin{abstract}
\noindent We describe the minimal configurations of the compact
$D=11$ Supermembrane and D-branes when the spatial part of the
world-volume is a K\"ahler manifold. The minima of the
corresponding hamiltonians arise at immersions into the target
space minimizing the K\"ahler volume. Minimal immersions of
particular K\"ahler manifolds into a given target space
are known to exist. They have
associated to them a symplectic matrix of central charges. We
reexpress the Hamiltonian of the $D=11$ Supermembrane with a
symplectic matrix of central charges, in the light cone gauge,
using the minimal immersions as backgrounds and the 
$Sp\parent{2g,\mathbb{Z}}$
symmetry of the resulting theory, $g$ being the genus of the
K\"ahler manifold. The resulting theory
is a symplectic noncommutative Yang-Mills theory coupled with the
scalar fields transverse to the Supermembrane. We prove that both
theories are exactly equivalent. A similar construction may be
performed for the Born-Infeld action. Finally, the noncommutative
formulation is used to show that the spectrum of the reguralized
Hamiltonian of the above mentioned $D=11$ Supermembrane is a
discrete set of eigenvalues with finite multiplicity.
\end{abstract}\thispagestyle{empty}\end{titlepage}

\newpage\section{Introduction} Recently, the spectrum of the
regularized compactified $D=11$
Supermembrane\hyphenation{Su-per-mem-bra-ne} \cite{Bergshoeff}
with non-trivial central charge, or equivalently the
Supermembrane with non-trivial wrapping on the compactified
directions of the target space, was shown to be a discrete set of
eigenvalues with finite multiplicity \cite{LB+MG+AR,LB+MG+IM+AR}
see also \cite{IM+AR+RT,Duff}. The spatial world-volume was
assumed to be a torus while the compactified part of the target
space was taken as $S^1\times S^1$. The regularization was
performed using the matrix model constructed in \cite{MG+AR}. It
was proven several\hyphenation{se-ve-ral} years ago that the
spectrum of the regularized $D=11$ Supermembrane in a Minkowski
target space is continuous from $\left[0,\infty\right)$
\cite{deWitt+Luscher+Nicolai,deWit+M+N}. Meanwhile the spectrum
of the compactified $D=11$ Supermembrane, including
configurations without non-trivial central charges, would be
continuous\hyphenation{con-ti-nuous} according to
\cite{deWitt+Peters+Plefka}, although there is no rigorous proof
in this case.

The Hamiltonian of the compactified $D=11$ Supermembrane with
non-trivial central charge has local minima for each winding of
the Supermembrane, characterized by an integer $n$, on the
compactified directions. When we consider a fixed $n$ the
one-forms $dX^r$, where $X^r$ are the maps from the world-volume
to the target space, may be decomposed into the
minima\hyphenation{mi-ni-ma} of the Hamiltonian $\hat{X}^r$ plus
an exact one form describing the degrees of freedom of the
Supermembrane with non-trivial central charge $n$. The resulting
action for those degrees of freedom was shown to be a symplectic
noncommutative Yang-Mills theory on the spatial world-volume $\mc{M}$
\cite{IM+JO+AR,IM+AR}. The equivalence between the Supermembrane
with non-trivial central charge and the symplectic noncommutative
Yang-Mills theory is exact, there is no approximation in the
construction.

In this paper we generalize this approach. We look for non-trivial
minima of the hamiltonians of the $D=11$ Supermembrane and
D-branes \cite{Polchinski}. Those are finite action solutions of
the field equations. We completely
characterize\hyphenation{cha-rac-te-ri-ze} those solutions when
$\mc{M}$ is a K\"ahler manifold. It was already shown in
\cite{B+R} that the extended self-dual configurations are finite
action solutions of the Born-Infeld theory, where the metric on
$\mc{M}$ is taken fixed and equal to the one constructed from the
two-form curvature and the almost complex structure in the usual
way. The minimal solutions correspond to particular solutions
saturating the Bogomolnyi bound.  States saturating this bound
were extensively used in the analysis of string dualities
\cite{Hull, Witten, Polchinski2}. In our analysis of the quantum
spectrum of the hamiltonian  of
the supermembrane we will use very specific properties of
the solutions which were derived mainly in the mathematical literature.
After the analysis of the minimal solutions we will consider the
Supermembrane. Once we have characterized the minimal solutions we will
consider the Supermembrane on this background and show that the
resulting theory is a symplectic noncommutative Yang-Mills theory
on the compactified directions of the target space which we will
consider to be a flat torus. We do so, because in these cases the
minimal\hyphenation{mi-ni-mal} solutions can be explicitly constructed. The central
charges associated to those solutions are expressed as a
symplectic matrix times the non-trivial winding of the membrane.
We also prove in those cases that the quantization of the
Supermembrane with a symplectic matrix of central charge is
exactly equivalent to the quantization of the symplectic
noncommutative theory and show that their regularized
hamiltonians have discrete spectrum. The analysis of the minimal
solutions for other compactifications like a $G_2$ holonomy
manifold\hyphenation{ma-ni-fold}, where $G_2$ is the automorphism
group of the octonions, which are very relevant in M-theory is
under investigation.

\section{Minimal solutions} The extended self-dual configurations
constructed in \cite{B+R} were defined on manifolds of even
dimension $2n$ where a curvature two-form satisfy the conditions
\[ ^*P_m \sim P_{n-m} \] for all possible $m$, where $P_m = F\wed
F\wed\ldots\wed F$ ($m$ factors) and $^*$ denotes the Hodge dual.
That is, the set $\llaves{P_m}$ is left invariant by the Hodge
dual operation. In four dimensions ($n=2$) it reduces to
$^*\parent{F\wed F}=\mathrm{constant}$ and $^*F\sim F$ the well
known self duality condition which solves the Maxwell equations.
In two dimensions $^*F=\mathrm{constant}$ represents a monopole
configuration\hyphenation{con-fi-gu-ra-tion}.

In \cite{B+R} it was shown that those configurations solve the
field equations of the Born-Infeld action. Moreover, it was
proven they are minima of the corresponding Hamiltonian.

In this paper we will consider the spatial world-volume $\mc{M}$
to be a K\"ahler manifold. The K\"ahler form is an extended
self-dual configuration with the corresponding K\"ahler metric, 
hence, at least as the field strength is
concern, it is a solution of the Born-Infeld field equation. This
is a good motivation to consider the Born-Infeld action over
K\"ahler manifolds. Also, most of the work on the $D=11$
Supermembrane has been performed for the case $\mc{M}$ is a
Riemann surface, which are K\"ahler manifold. We then restrict our
analysis to that case.

The Born-Infeld action of a D-brane \cite{Leigh} is a functional
of two fields over the D-brane world-volume $M$; they are the set
of components of a map $M\rightarrow N$, where $N$ is the target
space, and an $U\parent{1}$ gauge field $A$. The action is \beq
S\corchet{X,A} = \int\limits_M d^{p+1}\sigma\sqrt{-\det
E_{\hat{\mu}\hat{\nu}}} \label{covactionA}\eeq where \barr &&
E_{\hat{\mu}\hat{\nu}} = G_{\hat{\mu}\hat{\nu}} +
F_{\hat{\mu}\hat{\nu}} \\ &&
G_{\hat{\mu}\hat{\nu}}=\partial_{\hat{\mu}}X^{\hat{M}}
\partial_{\hat{\nu}}X^{\hat{N}}g_{\hat{M}\hat{N}}\parent{X} \\ &&
F_{\hat{\mu}\hat{\nu}}= \partial_{\hat{\mu}}A_{\hat{\nu}}-
\partial_{\hat{\nu}}A_{\hat{\mu}} \earr \[ \hat{M}=0,M
\h{,}\hspace{4mm} M=1,\ldots,\dim N-1\h{,}\hspace{4mm}
\hat{\mu}=0,\mu \h{,}\hspace{4mm}
\mu=1,\ldots,p\] $g_{\hat{M}\hat{N}}$ denotes a
metric on $N$ and $X^{\hat{M}}$ are the components of
the map $X$.

Equations of motion derived from \ec{covactionA} are \barr &&
\parent{E^{-1}}_+^{\hat{\nu}\hat{\mu}}
\pal_{\hat{\mu}}\pal_{\hat{\nu}} X^{\hat{M}} +\frac{1}{\sqrt{-\det
E_{\hat{\sig}\hat{\rho}}}}\pal_{\hat{\mu}}\corchet{\sqrt{-\det
E_{\hat{\sig}\hat{\rho}}}\parent{E^{-1}}_+^{\hat{\nu}\hat{\mu}}}
\pal_{\hat{\nu}} X^{\hat{M}} \nn\\ &&
+\parent{E^{-1}}_+^{\hat{\nu}\hat{\mu}}\pal_{\hat{\mu}}X^{\hat{L}}
\pal_{\hat{\nu}}X^{\hat{K}}\Gam^{\hat{M}}_{\hat{K}\hat{L}}=0\h{,}
\label{geneq1}\\ && \partial_{\hat{\mu}}\corchet{\sqrt{-\det
E_{\hat{\sig}\hat{\rho}}}\parent{E^{-1}}_-^{\hat{\mu}\hat{\nu}}}
=0 \label{geneq2} \earr where $\parent{E^{-1}}_+$ and
$\parent{E^{-1}}_-$ are the symmetric and antisymmetric parts of
$E^{-1}$ respectively and $\Gam^{\hat{M}}_{\hat{K}\hat{L}}$ are
the Christoffel symbols of the metric $g$ on the target. The first
(second) equation is obtained by taking variations of
\ec{covactionA} respect to $X^{\hat{M}}$ ($A_{\hat{\mu}}$).

We assume the foliation of the world-volume
$M=\mc{M}\times\mathbb{R}$, where $\mc{M}$ is an Riemannian
manifold. The first step in constructing our solutions is to impose
the following conditions on the fields
\barr && X^0=\sig^0 \h{,}\hspace{10mm} X^M=X^M\parent{\sig^\alp}
\label{static1} \\ && g_{00}=-1\h{,}\hspace{10mm} g_{0M}=0
\label{static3} \\ && A_0=0 \h{,}\hspace{13mm}
A_\mu=A_\mu\parent{\sig^\alp}\h{.} \label{static2}\earr The
resulting equations over $\mc{M}$ are \barr &&
\parent{E^{-1}}_+^{\nu\mu} \pal_\mu\pal_\nu X^M
+\frac{1}{\sqrt{\det E_{\sig\rho}}}\pal_\mu\corchet{\sqrt{\det
E_{\sig\rho}}\parent{E^{-1}}_+^{\nu\mu}}\pal_\nu X^M \nn \\ &&
+\parent{E^{-1}}_+^{\nu\mu}\pal_\mu X^L \pal_\nu
X^K\Gam^M_{KL}=0\h{,} \label{eq1} \\ &&
\partial_{\mu}\corchet{\sqrt{\det
E_{\sig\rho}}\parent{E^{-1}}_-^{\mu\nu}} =0 \h{.}\label{eq2}\earr
We also suppose the target space to be the product of a Minkowski
manifold times a compact Riemann manifold $\mc{N}$. For the
solution we fix those components of the map corresponding to the
Minkowski manifold equal to zero. Therefore, the remaining
degrees of freedom are $X^r$, $r=1,\ldots,\dim\mc{N}$, and
$A_\mu$, being both fields over $\mc{M}$. We consider $X^r$ to be
the components of an immersion $\mc{M}\rightarrow\mc{N}$, hence,
we assume that $G_{\mu\nu}=\partial_\mu X^r
\partial_\nu X^sg_{rs}$ is a Riemannian metric on $\mc{M}$.

Let $\mc{M}$ be a K\"ahler manifold of real dimension $p=2n$. Let
$J$, $K$ and $\Ome\parent{U,V}=K\parent{JU,V}$ be the almost
complex structure, the K\"ahler metric and the K\"ahler form of
$\mc{M}$ respectively. We suppose that $A_{\mu}$, $X^M$ and
$g_{MN}$ verify \beq G=K \h{,}\hspace{10mm}
F=\Ome\h{.}\label{ansatz}\eeq If \ec{ansatz} is verified, the
Born-Infeld tensor becomes $E_{\sig\rho}=\parent{\del_\sig^\alp+
J_\sig^\alp}K_{\alp\rho}$ and then we get the remarkable results
\barr && \det E_{\sig\rho}=2^n\det K_{\sig\rho} \label{deter}\\ &&
\parent{E^{-1}}_+^{\mu\nu}={1\over 2}K^{\mu\nu}
\h{,}\hspace{10mm}\parent{E^{-1}}_-^{\mu\nu} = -{1\over
2}\Ome^{\mu\nu}\label{inverse}\earr where Greek indices are
raised and lowered with the K\"ahler metric.

Note that eqs.
\ec{eq2} are automatically solved by \ec{ansatz}. The fact that the K\"ahler
geometry ans\"atz \ec{ansatz} solves the Born-Infeld equation
\ec{eq2} is a particular\hyphenation{par-ti-cu-lar} case of the most general result about
extended self-dual configurations, as we mentioned previously. 
Moreover, the K\"ahler form of any K\"ahler
manifold may be locally expressed in terms of the K\"ahler
potential $\mc{K}$, $\Ome=\im\partial\bar{\partial}\mc{K}$ where
$\pal$ and $\bar{\pal}$ are Dolbeault operators. This ensures the
local existence of an one-form \beq \alp\equiv{\im\over
2}\parent{\bar{\partial}-\partial}\mc{K} \eeq which locally
verifies $\Ome=d\alp$, hence $\alp$ may be taken as $A$.

Under \ec{ansatz}, equations \ec{eq1} yields \beq
K^{\nu\mu}\pal_\mu\pal_\nu X^r -K^{\mu\nu} \pal_\sig
X^r\gam^\sig_{\mu\nu} +K^{\nu\mu}\pal_\mu X^t \pal_\nu
X^u\Gam^r_{tu}=0 \label{fineq} \eeq where $\gam$ is the
Christoffel symbol of the K\"ahler metric $K$. The first two
terms form the Laplacian defined with the K\"ahler metric for
each component $X^r$. It is straightforward to show that \ec{eq1}
and \ec{eq2} are exactly the same equations to be satisfied by the
minima of the Hamiltonian of the Born-Infeld action. Hence,
\ec{ansatz} and \ec{fineq} describe the minima of the Hamiltonian
over a K\"ahler spatial world-volume.

Equation \ec{fineq} can be derived
from the volume action \[ \int\limits_{\mc{M}} d^p\sig\sqrt{\det
K_{\sig\rho}} \] where \[ K_{\mu\nu}=\pal_\mu X^r \pal_\nu X^s
g_{rs}\h{.}\] Therefore, the solutions for the remaining equation
\ec{fineq} represents actually\hyphenation{ac-tua-lly}
minimal immersions, in the mathematical
nomenclature, of the K\"ahler manifold in the target space.

The condition of being an immersion is a nontrivial one. The
search for minimal immersions, which in term are directly related
to harmonic maps between Riemannian manifolds, has been followed
in the mathematical literature \cite{Eells+Sampson}. Interesting
results has been obtained when the target space is a flat torus.
In this case minimal immersions of certain K\"ahler manifolds are
known to exist. An explicit example is the Albanese map of a
compact K\"ahler manifold $\mc{M}$ whose cotangent bundle is ample
\cite{Matsushima,Nagano}. In particular the Albanese map of a
Riemannian surface of positive genus into a flat torus is a
minimal immersion. This minimal immersion is the one we will
consider in the construction of a noncommutative gauge theory.

\section{Noncommutative Yang-Mills from compactified $D=11$ membranes}
In this section we will show that using the minimal immersion as a
background we can reexpress the Born-Infeld action or the $D=11$
Supermembrane action as a symplectic noncommutative Yang-Mills
theory. We will perform explicit calculations for the $D=11$
Supermembrane, although the same arguments follow directly for
the Born-Infeld action.

We consider the Hamiltonian of the $D=11$ Supermembrane in the
light cone gauge \cite{deWitt,Ovalle}. Its bosonic part is given
by \beq \mc{H}=
\frac{1}{2\sqrt{W}}\corchet{P_mP^m+\det\parent{\pal_\mu
  X^m\pal_\nu X_m}}\label{smham}\eeq
subject to the area preserving diffeomorphisms constraints \[
\eps^{\mu\nu}\pal_\mu\parent{\frac{1}{\sqrt{W}} P_m\pal_\nu X^m}
= 0 \] where $X^m$, $m=1,\ldots,9$ are maps from $\mc{M}$ a
Riemannian surface of genus $g$ to a flat target space. $P_m$ are
its conjugate momenta, the transverse indices $m$ are raised and
lowered with the flat metric. $\sqrt{W}$ is a scalar density
introduced in the Light Cone Gauge fixing procedure.

The local isolated minima of the Hamiltonian $\mc{H}$ are
immersions satisfying \[ P_m =0 \] and
\[ G^{\nu\mu}\pal_\mu\pal_\nu X^m -G^{\mu\nu}
\pal_\sig X^m\gam^\sig_{\mu\nu} +G^{\nu\mu}\pal_\mu X^l \pal_\nu
X^k\Gam^m_{kl}=0 \] the same equations \ec{fineq}, the third term
being zero in the case we
consider a flat target space. In this particular case this equation,
when rewritten in terms of a local complex coordinate $z$ and its
complex conjugate, reduces to \[ \pal_z\pal_{\bar{z}} X^m=0 \h{.} \]
Hence, the minimal immersions are harmonic maps. The condition of
having an immersion is in any case a nontrivial one.

We will assume that the target space is compactified and that the
Supermembrane has non-trivial winding over it. Although several
of our arguments are general we will specifically consider
Supermembranes with non-trivial winding over $M_9\times S^1\times
S^1$ and $M_7\times S^1\times S^1\times S^1\times S^1$, where
$M_9$ and $M_7$ are Minkowski spaces while $S^1\times S^1$ and
$S^1\times S^1\times S^1\times S^1$ are flat torus $T$. The case
$M_9\times S^1\times S^1$ was considered extensively in
\cite{IM+AR+RT,IM+JO+AR,LB+MG+IM+AR,LB+MG+AR}. A rigorous proof
was given showing that the spectrum of the $D=11$ Supermembrane
with a fixed non-trivial central charge is discrete. The analysis
for more general target spaces requires the introduction of a
minimal immersion of $\mc{M}$ onto $T$. This ingredient of the
construction becomes straightforward in the case $T$ is $S^1\times
S^1$, and is implicit in the previous works. However, it becomes
a nontrivial step in the construction for more general
compactified target spaces.

The condition describing a map from $\mc{M}$ to $M\times T$ is
expressed by \beq \oint\limits_{\mc{C}} dX^r=m^r \label{torus}\eeq
where $m^r$ are integers, while $\mc{C}$ represents a basis of
homology over $\mc{M}$. $X^r$ denote the compactified directions
on the target space. The topological index describing the
condition of nontrivial winding is the integral over $\mc{M}$ of
the pull-back of the symplectic form on the torus $T$: \beq
\int\limits_\Sig dX^r\wed dX^s = 2\pi n \ome^{rs}
\label{winding}\eeq where $\ome^{rs}$ is the canonical symplectic
matrix on $T$. The area of $\mc{M}$ has been normalized to $2\pi$.
We will consider all maps satisfying conditions \ec{torus} and
\ec{winding}. They ensure the membrane has non-trivial central
charges.

We know, from the previous section, the existence of a minimal
immersion from any compact Riemann surface of genus $g>0$ to the
corresponding Jacobian variety. This map is also known in the
literature as the Albanese map. In order to be specific, we will
consider $\mc{M}$ to be a Riemann surface of genus 2 such that its
Jacobian variety is $S^1\times S^1\times S^1\times S^1$. The
minimal immersion is obtained from the harmonic one-forms
$d\hat{X}^r$, $r=1,\ldots,4$, over $\mc{M}$ satisfying condition
\ec{winding}. They have the property that \[
\pal_\mu\hat{X}^r\pal_\nu\hat{X}^s\ome_{rs} d\sig^\mu \wed
d\sig^\nu \] where $\sig^\mu$, $\mu=1,2$, are local coordinates on
$\mc{M}$, is the K\"ahler two-form $\Ome$ over $\mc{M}$ satisfying
\beq ^*\Ome =n \label{dual} \eeq where the Hodge dual is
constructed with the K\"ahler metric \[ K_{\mu\nu} =
\pal_\mu\hat{X}^r\pal_\nu\hat{X}^s\del_{rs} \h{.}\]

We may now consider a general map from $\mc{M}$ to $T$, it can
always be decomposed as \beq dX^r =
m^r_{s}d\hat{X}^s+\del^{rs}d\mc{A}_s\h{,} \label{prechange}\eeq
where as before $d\hat{X}^s$, $s=1,\ldots,4$, denotes a basis of
harmonic one-forms realizing the minimal immersion from $\mc{M}$
to $T$ while $d\mc{A}_s$ are exact one forms. $m^r_s$ must be
integers in order to satisfy conditions \ec{torus}. Moreover,
from \ec{winding}, they must satisfy \[m^t_rm^u_s\ome_{tu} =
\ome_{rs} \h{,} \] hence the matrices with elements $m^r_s$
belong to $Sp\parent{4,\mathbb{Z}}$ the modular group for a
Riemann surface of genus 2. We notice from \ec{dual} that the
diffeomorphisms over $\mc{M}$ which change the basis of harmonic
one-forms by elements of $Sp\parent{4,\mathbb{Z}}$ preserve
$\sqrt{W}$ since \[ \eps^{\mu\nu}
\pal_\mu\hat{X}^r\pal_\nu\hat{X}^s \ome_{rs} = n\sqrt{W} \h{,}
\]  where $W$ has been identified with the determinant of the
K\"ahler metric over $\mc{M}$, hence those diffeomorphisms are
gauge symmetries of the theory. We may then consider in
\ec{prechange} a fixed basis of harmonic one forms satisfying
condition \ec{winding}. That is we can always consider \beq dX^r
= d\hat{X}^r+\del^{rs}d\mc{A}_s\h{.} \label{change}\eeq The
degrees of freedom of the $D=11$ membrane satisfying the
topological condition \ec{winding} are then described by the
exact one-forms $d\mc{A}_r\parent{\sig}$ and $dX^m$,
$r=1,\ldots,4$, $m=1,\ldots,5$, modulo the area preserving gauge
transformations.

We now replace \ec{change} into the Hamiltonian \ec{smham}. The
action of the Supermembrane is then exactly expressed as the
value of the Supermembrane action at the minimal immersion, a
finite action solution of the field equations plus a positive
Hamiltonian for the $\mc{A}_r$ and $X^m$, $r=1,\ldots,4$,
$m=1,\ldots,5$, fields. It turns out that this is the pull-back
to $\Sig$ of a symplectic noncommutative Yang-Mills theory in
$M\times T$.

We first note that the kinetic terms $P\dot{X}$ may be rewritten as \[
P_r\dot{X}^r + P_m\dot{X}^m =
\parent{\del^{rs}P_s}\dot{\mc{A}}_r+P_m\dot{X}^m \] since $\hat{X}^r$
do not depend on $\tau$. We will denote $\pi^r\equiv \del^{rs}P_s$
the conjugate momenta to $\mc{A}_r$. We then introduce the
derivatives \[\mc{D}_r \equiv D_r + \llaves{\mc{A}_r,\;}\] where
\basn && D_r=e_r^\mu\pal_\mu \h{,} \\ && e^\mu_r \equiv
\Ome^{\mu\nu}\pal_\nu\hat{X}^s\del_{sr}\easn and \[
\llaves{\phi,\varphi}= \Ome^{\mu\nu} \pal_\mu\phi\pal_\nu\varphi =
\frac{2}{n}\ome^{rs}D_r\phi D_s\varphi \h{.}\] The symplectic
noncommutative curvature is then given by \[ \mc{F}_{rs} =
D_r\mc{A}_s - D_s\mc{A}_r + \llaves{\mc{A}_r,\mc{A}_s} \h{.}
\] The final form of the bosonic Hamiltonian density, in terms of
the above geometrical objects, is \basn \mc{H} &=&
\frac{1}{2\sqrt{W}} \left(P_mP^m+ \pi^r\pi_r +
W\mc{D}_rX^m\mc{D}^rX_m+ \frac{1}{2}W\mc{F}_{rs}\mc{F}^{rs}\right. \\ 
& &\left.+
\frac{1}{2}W\llaves{X^m,X^n}^2\right)+\frac{1}{8}\sqrt{W}n^2+ \lam\parent{
\mc{D}_r\pi^r+\llaves{X^m,P_m}} \easn where indices are raised
and lowered with the corresponding flat metrics. $\lam$ is the
Lagrange multiplier associated to the volume preserving
constraint\hyphenation{cons-traint} which is now interpreted as
the Gauss constraint for the noncommutative formulation of the
$D=11$ Supermembrane with non-trivial fixed central charge.

The above Hamiltonian density has the same expressions as the one
analyzed in \cite{LB+MG+IM+AR} although the range of indices are
different. Following \cite{LB+MG+IM+AR} one may show: i) The
non-existence of string-like spikes for $\mc{H}$. ii) The
regularized Hamiltonian has a potential bounded below and becoming
infinite at infinity in every possible direction on the
configuration space. Consequently, using lemma 1 and the results
of section 5 and the general criteria discussed in the conclusions
of \cite{LB+MG+AR}, one can show that the spectrum of the
complete quantum regularized Hamiltonian, including the fermionic
terms, is a discrete set of eigenvalues with finite multiplicity.

If we now integrate out the conjugate momenta $P_m$ and $\pi^r$
from the corresponding functional integral we obtain: \[ \mc{L} =
-\sqrt{W} \corchet{\frac{1}{2}\parent{\mc{D}_{\hat{r}}X^m}^2 +
\frac{1}{4}\parent{\mc{F}_{\hat{r}\hat{s}}}^2+
\frac{1}{4}\llaves{X^m,X^n}^2}-\frac{1}{8}\sqrt{W}n^2 \]
\[ \hat{r}=0,r\h{,}\hspace{10mm}r=1,\ldots,4\] where \basn && \mc{A}_0
\equiv \lam \\ && \mc{D}_0X^m= \dot{X}^m+\llaves{\mc{A}_0,X^m} \\
&& \mc{F}_{0r} = \dot{\mc{A}_r} - D_r\mc{A}_0 +
\llaves{\mc{A}_0,\mc{A}_r} \easn and $\mc{D}_r$, $\mc{F}_{rs}$ are
defined as before. The final expression of the Lagrangian is the
pull-back to $\mc{M}$ of a noncommutative Yang-Mills in the
space-time constructed with the light cone $\tau$ and the
compactified subspace of the target space.

\section{Conclusions} We described the minimal configurations of the
compact Supermembrane and D-brane theories, when the spatial part
of the world-volume $\mc{M}$ is a K\"ahler manifold. We showed
that the minima of their hamiltonians take place at immersions
from $\mc{M}$ into the target space minimizing the K\"ahler
volume. The explicit construction of some minimal immersions of
K\"ahler manifold with ample cotangent bundle has been described
in the mathematical\hyphenation{ma-the-ma-ti-cal} literature. We
follow that construction for the case the target space is the
product of a Minkowski space times a flat torus. We prove that a
$D=11$ Supermembrane with non-trivial central charges on a target
space of that kind is exactly equivalent to the pullback of a
symplectic noncommutative Yang-Mills theory over the compactified
sector of the target space coupled to the transverse scalar
fields. The construction makes explicit use of the minimal
immersion which is taken as background to reexpress the
Supermembrane action as a symplectic noncommutative one. We
emphasize that there is no approximation in the construction. We
expect the same result for any target space of the form $M\times
T$ provided there exists a minimal immersion from $\mc{M}$ into
$T$.

The analysis was performed for the bosonic sector of the
Supermembrane. The addition of the fermionic terms does not
change any of the results. Moreover, applying the propositions
proven in \cite{LB+MG+AR} we conclude that the spectrum of the
regularized $D=11$ Supermembrane with target space $M\times T$,
$T=S^1\times S^1\times S^1\times S^1$ and with a symplectic
matrix of central charges given by eq. \ec{winding} consists of a
discrete set of eigenvalues with finite multiplicity.

\end{document}